\documentclass[aps,amsfonts,prd,twocolumn]{revtex4}

\usepackage{amsfonts}
\usepackage{amsmath}
\usepackage{amssymb}
\usepackage{epsfig}
\usepackage{graphicx}

\newcommand{\bs}{\boldsymbol}
\newcommand{\ar}{\arrowvert}
\newcommand{\ra}{\rangle}

\newcommand{\be}{\begin{equation}}
\newcommand{\ee}{\end{equation}}
\newcommand{\ba}{\begin{eqnarray}}
\newcommand{\ea}{\end{eqnarray}}

\begin{document}

\title{QCD Coulomb gauge approach to hybrid mesons}

\author{Ignacio J. General} 
\affiliation{Department of Physics, North Carolina State
University, Raleigh, North Carolina 27695-8202}

\author{Felipe J. Llanes-Estrada} 
\affiliation{Departamento de Fisica Teorica I, Universidad Complutense
28040 Madrid, Spain}

\author{Stephen R. Cotanch} 
\affiliation{Department of Physics, North Carolina State University,
Raleigh, North Carolina 27695-8202}

\date{\today}

\begin{abstract}
An effective Coulomb gauge Hamiltonian, $H_{\rm eff}$, is used to
calculate the light ($u\bar{u}g$), strange ($s\bar{s}g$) and
charmed ($c\bar{c}g$)  hybrid meson spectra. For the same two
parameter  $H_{\rm eff}$ providing glueball masses consistent with
lattice results and a good description of the observed
$u,d,s$ and $c$ quark mesons,  a large-scale variational treatment predicts the
lightest hybrid has $J^{PC} = 0^{++}$  and  mass 2100 MeV.
The lightest exotic $ 1^{-+}$ state is just above 2200
MeV, near  the upper limit of lattice and Flux Tube predictions.
These theoretical formulations all indicate  the observed $ 1^{-+}$
$\pi_1(1600)$ and, more clearly, $\pi_1(1400)$ are not hybrid
states. The Coulomb gauge approach further predicts that in the strange and charmed sectors, respectively,
the ground
state hybrids  have $1^{+-}$ with masses
2125 and 3830 MeV, while the first exotic
$1^{-+}$  states are at 2395 and 4020 MeV.
Finally, using our hybrid wavefunctions, dimensional counting rules and the Franck-Condon principle, novel experimental
signatures are presented to assist light and heavy hybrid meson searches.
\end{abstract}

\pacs{12.39.Mk; 12.39.Pn; 12.39Ki; 12.40.Yx}
\keywords{Hybrid mesons; Exotic states; Effective Hamiltonian; QCD Coulomb gauge;}

\maketitle

\section{Introduction}

Following  the ``Eightfold Way" (Gell-Mann and Ne'eman), the
``Quark Model" (Gell-Mann and Zweig), along with subsequent
extensions, has generally explained  the observed hadronic spectrum.
This is especially true for heavy flavored mesons where it is now clear that higher
order QCD corrections can be ignored or treated perturbatively. Even
in the light sector, the phenomenological quark model works
reasonably well. However, the existence of hadrons with exotic
quantum numbers (i.e. $J^{PC}$ states not possible in  $q \bar{q}$
or $qqq$ systems) clearly reveals this model is not complete.
Related, it is  expected that there are exotic hadrons with
conventional quantum numbers that also can not be described by the
quark model, e.g., glueballs $gg$, hybrid mesons $q\overline{q}g$
and tetraquarks $q\overline{q}q\overline{q}$.

Possible experimental evidence for a $1^{-+}$ exotic state was
first reported in 1988 \cite{Alde}, but the situation was not
clarified until several years later.  Now it is believed
that there exist two states with these quantum numbers
below 2 GeV: $\pi_1(1400)$ \cite{E852-1,E852-2} and $\pi_1(1600)$
\cite{E852-3,E852-4}(note, a recent analysis~\cite{ds} finds no
evidence for either candidate).
There are also other reported hybrid
candidates with $J^{PC} =$ $0^{-+}$ \cite{Amelin,Zaitsev},
$1^{--}$ \cite{Donnachie} and
$2^{-+}$ \cite{Karch,Adomeit,Barberis}.

Theoretically, the structure of the $\pi_1$ states remains
unclear. They could be hybrid or tetraquark mesons with most
theoretical studies \cite{Kalashnikova:2001ke,Buisseret:2006sz}  investigating the former. Lattice gauge
simulations \cite{Bernard1,Bernard2,Lacock,Hedditch,Luo} predict
the lightest hybrid meson is between 1.7 and 2.1 GeV and results
from the  Flux Tube model \cite{Barnes,Close,Katja} also
span much of this range. Only vintage Bag model \cite{Bag}
calculations yield a  lower mass, between 1.3 and 1.8 GeV, but
Ref. ~\cite{Iddir} argues that the $\pi_1(1400)$ is not a hybrid.
Table \ref{table:comparison-u} list predictions for the
$u/d$, $s$ and $c$ $1^{-+}$ hybrid mesons.

\begin{table} [h]
\begin{ruledtabular}
  \begin{tabular}{|c|c|c|c|}
Model [Reference] &    $u/d$ hybrid & $s$ hybrid & $c$ hybrid   \\
    \hline
    \hline
Lattice QCD [15-19, 25-27]&   1.7 - 2.1 & 1.9 &  4.2 - 4.4   \\
Flux Tube  \cite{Barnes,Close,Katja} &    1.8 - 2.1 & 2.1 - 2.3  & 4.1 - 4.5 \\
Bag Model   \cite{Bag} &    1.3 - 1.8 & & 3.9  \\
  \end{tabular}
  \caption{Published predicted  $1^{-+}$  masses, in GeV, for light, strange and charmed hybrid mesons.}
  \label{table:comparison-u}
\end{ruledtabular}
\end{table}

In this work, we study $q\bar{q}g$ hybrid states using a field
theoretical, relativistic many-body approach based upon an
effective QCD Hamiltonian, $H_{\rm eff}$, formulated in the
Coulomb gauge. This model  successfully describes
the meson spectrum \cite{LC2,LCSS} and is also  consistent
\cite{LBC} with lattice  glueball (and oddball) predictions. Using
standard bare current  quark masses, it properly incorporates
chiral symmetry, yet dynamically generates a constituent mass and
spontaneous chiral symmetry breaking \cite{LC1}. Further, it
provides a good description of the vacuum properties (quark and
gluon condensates),  and
respects the global, internal symmetries
of QCD, as well as the spatial Euclidean group,
all
within a minimal two parameter theory. Our work also extends an
earlier hybrid calculation \cite{LChybrid} by including previously omitted
terms in the Hamiltonian and  by
comprehensively predicting the light, strange and charmed  hybrid
meson spectra.

This paper is organized into 8 sections. In Sections II and III the effective
Hamiltonian is presented along with
an improved hyperfine interaction which
provides  realistic spin splittings in both light and heavy mesons
\cite{LCSS} and, for the first
time, the rigorous non-abelian contributions from the color
magnetic fields. Section IV details the corresponding improved quark and
gluon gap equations  and a variational formulation
for the hybrid meson problem is developed in Sec. V.  Calculations and new
results are discussed in Sec. VI, while in Sec. VII we develop novel
experimental signatures for observing hybrid mesons having both conventional and exotic
quantum numbers.
Finally, we summarize  results and conclusions in Sec. VIII.

\section{Effective Hamiltonian}

Our effective, quark-gluon Hamiltonian is an approximation to the
exact Coulomb gauge QCD Hamiltonian  \cite{T-D-Lee} and  is given
by (summation over repeated indices is used throughout this
paper)
\begin{eqnarray}
H_{\rm eff} &=& H_q + H_g +H_{qg} + H_{C}   \\
H_q &=& \int d{\bs x} \Psi^\dagger ({\bs x}) [ -i {\mbox{\boldmath$\alpha$\unboldmath}} \cdot {\mbox{\boldmath$\nabla$\unboldmath}}
+  \beta m] \Psi ({\bs x})   \\
H_g &=& \frac{1}{2} \int d {\bs x}\left[ {\bf \Pi}^a({\bs x})\cdot {\bf
\Pi}^a({\bs x}) +{\bf B}^a({\bs x})\cdot{\bf B}^a({\bs x}) \right] \\
H_{qg} &=&  g \int d {\bs x} \; {\bf J}^a ({\bs x})
\cdot {\bf A}^a({\bs x}) \\
H_C &=& -\frac{1}{2} \int d{\bs x} d{\bs y} \rho^a ({\bs x}) \hat{V}(\ar {\bs x}-{\bs y}
\ar ) \rho^a ({\bs y})   \ .
 \label{model}
\end{eqnarray}
Here $g$ is the QCD coupling, $\Psi$ is the quark field with
current quark mass $m$, ${\bf A}^a$ are the gluon fields
satisfying the transverse gauge condition,
$\mbox{\boldmath$\nabla$\unboldmath}$ $\cdot$ ${\bf A}^a = 0$, $a
= 1, 2, ... 8$, ${\bf \Pi}^a $ are the conjugate fields and ${\bf
B}^a$ are the non-abelian magnetic fields
\begin{eqnarray}
{\bf B}^a = \nabla \times {\bf A}^a + \frac{1}{2} g f^{abc} {\bf A}^b \times {\bf A}^c \ .
\end{eqnarray}
The color densities, $\rho^a({\bs x})$, and quark color currents, ${\bf J}^a$, are
related to the fields by
\begin{eqnarray}
\rho^a({\bs x}) &=& \Psi^\dagger({\bs x}) T^a\Psi({\bs x}) +f^{abc}{\bf
A}^b({\bs x})\cdot{\bf \Pi}^c({\bs x}) \\
{\bf J}^a &=& \Psi^\dagger ({\bs x}) \mbox{\boldmath$\alpha$\unboldmath}T^a \Psi ({\bs x})
\ ,
\end{eqnarray}
where
$T^a = \frac{\lambda^a}{2}$ and $f^{abc}$ are  the
$SU_3$ color matrices and structure constants, respectively.

The bare parton fields have the following normal mode expansions
(bare quark spinors $u, v$, helicity, $\lambda = \pm 1$, and
color vectors $\hat{\boldsymbol{\epsilon}}_{{\cal C }= 1,2,3}$)
\begin{eqnarray}
\label{colorfields1}
 \Psi(\boldsymbol{x}) &=&\int \!\! \frac{d
    \boldsymbol{k}}{(2\pi)^3} \Psi_{{\cal C}} (\boldsymbol{k})  e^{i \boldsymbol{k} \cdot \boldsymbol{x}} \hat{\boldsymbol{\epsilon}}_{\cal C}  \\
\Psi_{ {\cal C}} (\boldsymbol{k})   & = & {u}_{\lambda} (\boldsymbol{k}) b_{\lambda {\cal C}}(\boldsymbol{k)}  + {v}_{\lambda} (-\boldsymbol{k})
    d^\dag_{\lambda {\cal C}}(\boldsymbol{-k)}   \\
{\bf A}^a({\bs{x}}) &=&  \int
\frac{d{\bs{k}}}{(2\pi)^3}
\frac{1}{\sqrt{2k}}[{\bf a}^a({\bs{k}}) + {\bf a}^{a\dag}(-{\bs{k}})]
e^{i{\bs{k}}\cdot
{\bs {x}}}  \ \ \
\\
{\bf \Pi}^a({\bs{x}}) &=& \hspace{-.15cm}-i \int \!\!\frac{d{\bs{k}}}{(2\pi)^3}
\sqrt{\frac{k}{2}}
[{\bf a}^a({\bs{k}})-{\bf a}^{a\dag}(-{\bs{k}})]e^{i{\bs{k}}\cdot
{\bs{x}}}  \!,
\end{eqnarray}
with the Coulomb gauge transverse condition, ${\bs k}\cdot {\bf
a}^a ({\bs k}) = \\ (-1)^\mu k_{\mu} a_{-\mu} ^a ({\bs k}) =0$.
Here $b_{\lambda {\cal C}}(\boldsymbol{k)}$, $d_{\lambda {\cal
C}}(\boldsymbol{-k)} $ and  $a_{\mu}^a({\bs{k}})$ ($\mu = 0, \pm
1$) are the bare quark, anti-quark and gluon Fock operators, the latter
satisfying the  transverse commutation relations,
\begin{equation}
[a^a_{\mu}({\bs k}),a^{b \dagger}_{\mu'}({\bs k}')]=
(2\pi)^3 \delta_{ab} \delta^3({\bs k}-{\bs k}')D_{{\mu} {\mu'}}({\bs k})  \ ,
\end{equation}
with
\begin{equation}
D_{{\mu} {\mu'}}({\bs k}) =
\delta_{{\mu}{\mu'}}- (-1)^{\mu}\frac{k_{\mu} k_{-\mu'}}{k^2}  \  .
\end{equation}

Confinement is described by a Cornell type potential,

\ba
\label{2}
\hat{V} (r = |{\bs x} - {\bs y}|) &=& {\hat V}_C (r) + {\hat V}_L (r) \\
    \hat{V}_C(r) &=& -\frac{\alpha_s}{r} \\
 \hat {V}_L(r)   &=& \sigma r ,
\ea
where  the string tension,  $\sigma=0.135$ GeV$^{2}$, and
$\alpha_s=\frac{g^2}{4\pi}=0.4$ have been previously determined.
The Fourier transform of $\hat V$ is denoted by $V$ and in momentum space these potentials  take the form $V_L(|\boldsymbol{p}|) = - 8 \pi \sigma /p^{4}$, $V_C(|\boldsymbol{p}|) = - 4 \pi \alpha_s /p^{2}$.
For comparison and also to provide hadronic structure sensitivity to potential form,
we report predictions using a confining potential~\cite{SS} having a renormalization
improved short-ranged behavior.  This potential was utilized in a previous meson
study~\cite{LCSS} and has the  momentum space representation
\begin{equation}
    { V}(|\boldsymbol{p}|) = \left\{ \begin{array}{ll}
        -\frac{8.04}{p^2}
    \frac{\ln^{-0.62}(\frac{p^2}{m_g^2}+0.82)}{\ln^{0.8}(\frac{p^2}{m_g^2}+1.41)}& \textrm{ $p>m_g$}\\
        -\frac{12.25 m^{1.93}_g}{p^{3.93}} & \textrm{$p<m_g$}
        \end{array} \right.  \ .
\label{adampot}
\end{equation}
The parameter $m_g$ sets the string tension and is  related to  $\sigma$ by
$m_g \cong \sqrt{ 8 \pi \sigma/12.25}\approx 600$ MeV.

\section{Hamiltonian $g^2$ Corrections}

As mentioned above, a previous hybrid application \cite{LChybrid}
used this Hamiltonian but set the QCD coupling, $g$, to zero. This
truncation eliminated the quark-gluon interaction, ${\bf J}^a
\cdot {\bf A}^a$, or ``hyperfine" term, Eq. (4), and also the non-linear
(non-abelian) component of the color magnetic fields, Eqs. (3, 6). Now, both
are included so that the  non-confining part of
the Hamiltonian is consistent to order $g^2$.

\subsection{Hyperfine correction}

\begin{figure} [t]
    \includegraphics[width=0.5\textwidth]{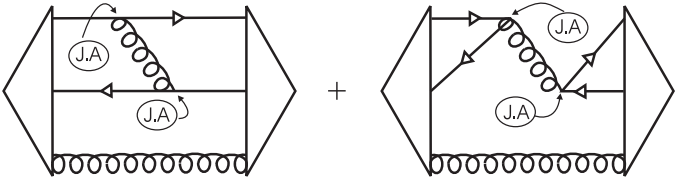}
    \caption{The hyperfine correction entails the exchange of a gluon between $q$ and
    $\overline{q}$ and $q\overline{q}$ annihilation.}
    \label{fig:JJdiagram}
\end{figure}

Following \cite{LCSS}, the Hamiltonian term $H_{qg}$ containing
the ${\bf J}^a \cdot {\bf A}^a$ operators is included using
perturbation theory to second order in $g$. Then, integrating over
the gluonic degrees of freedom yields an effective quark hyperfine
interaction with a ${\bf J}^a \cdot {\bf J}^a$ form. This contribution is
represented by the Feynman diagrams in Fig. \ref{fig:JJdiagram}.

The resulting transverse hyperfine interaction is
\begin{equation} \label{JJdiagram}
V_{T} =\frac{1}{2}\int \!\!\! \int \!\!
d\boldsymbol{x}d\boldsymbol{y}
    J_i^a(\boldsymbol{x}) \hat{U}_{i j}(\boldsymbol{x,y}) J_j^a(\boldsymbol{y}),
\end{equation}
where the kernel reflects the transverse gauge
\begin{equation}
    \hat{U}_{i j}(\boldsymbol{x,y}) = \bigg(\delta_{i j}- \frac{\nabla_i
    \nabla_j}{\nabla^2}\bigg)_{\boldsymbol{x}}\hat{U}(|\boldsymbol{x-y}|).
\end{equation}
For $\hat{U}$ we choose a modified Yukawa potential which incorporates a
dynamical mass, $m_g = 600$ MeV, for the exchanged gluon as
explained in \cite{LCSS}. Fourier transforming to momentum space,
this continuous potential takes the form
\begin{equation} \label{Adam's potential}
    { U}(p) = \left\{ \begin{array}{ll}
        -\frac{8.04}{p^2}
    \frac{\ln^{-0.62}(\frac{p^2}{m_g^2}+0.82)}{\ln^{0.8}(\frac{p^2}{m_g^2}+1.41)}& \textrm{ $p>m_g$}\\
        -\frac{24.50}{p^2+m_g^2} & \textrm{$p<m_g$}
        \end{array} \right. \ .
\end{equation}

\begin{figure} [h]
\includegraphics[width=0.3\textwidth]{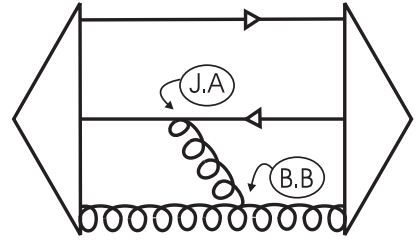}
  \caption{The non-abelian correction with triple gluon vertices.}
  \label{fig:3Gdiagram}
\end{figure}

\subsection{Non-abelian correction}

Similarly, the non-abelian components of the color magnetic
fields, $g f^{abc} {\bf A}^b \times {\bf A}^c$, in the kinetic energy are also included perturbatively
through
$({\bf J}^a\cdot {\bf A}^a)({\bf B}^a\cdot {\bf B}^a)$.  The resulting non-abelian interaction  is
represented by the Feynman diagram in Fig. \ref{fig:3Gdiagram}
and
is given by

\begin{widetext}
  \begin{eqnarray}
    \lefteqn{V_{NA}=\frac{1}{4}\int \!\!\! \int \!\! d\boldsymbol{x} d\boldsymbol{y} f^{abc} \varepsilon_{ijk}
    \varepsilon_{ilm} [J^a_h(\boldsymbol{y})
    (\nabla_j^{(\boldsymbol{x})} \hat{U}_{k h}(\boldsymbol{x,y}))
    A^b_l(\boldsymbol{x}) A^c_m(\boldsymbol{x}) +} \nonumber \\
    && J^b_h(\boldsymbol{y})
    (\nabla_j^{(\boldsymbol{x})} A^a_k(\boldsymbol{x})) \hat{U}_{l h}(\boldsymbol{x,y})
    A^c_m(\boldsymbol{x}) + J^c_h(\boldsymbol{y})
    (\nabla_j^{(\boldsymbol{x})} A^a_k(\boldsymbol{x})) A^b_l(\boldsymbol{x})
    \hat{U}_{m h}(\boldsymbol{x,y})].
  \end{eqnarray}
\end{widetext}
where $\hat{U}_{ij}$ is the same kernel appearing in the hyperfine
potential.

\section{Gap Equation}

Having defined the model Hamiltonian, the next step is to calculate
the ground state. Since we are free to expand the field operators
in any complete basis, we follow the Bardeen-Cooper-Schriffer (BCS) method
and perform a Bogoliubov-Valatin rotation,
\begin{eqnarray} \label{eq:operator rotations}
    B_{\lambda {\cal C}}(\boldsymbol{k)} &=& \cos\frac{\theta_k}{2}
    b_{\lambda {\cal C}}(\boldsymbol{k)}  - \lambda \sin\frac{\theta_k}{2}
    d^\dag_{\lambda {\cal C}}(\boldsymbol{-k)}   \nonumber  \\ \nonumber
    D_{\lambda {\cal C}}(\boldsymbol{-k)}&=& \cos\frac{\theta_k}{2}
    d_{\lambda {\cal C}}(\boldsymbol{-k)}  + \lambda \sin\frac{\theta_k}{2}
    b^\dag_{\lambda {\cal C}}(\boldsymbol{k)}  \\
    {\boldsymbol \alpha}^a(\boldsymbol{k}) &=& \cosh \Theta_k {\bf a}^a(\boldsymbol{k}) + \sinh \Theta_k
    {{\bf a}^a}^\dag(-\boldsymbol{k}) \ ,
\end{eqnarray}
which transforms the bare particle operators ${\bf a}^a$, $b_{\lambda {\cal C}}$ and $d_{\lambda {\cal C}}$ into the dressed,
quasi-particle operators ${\boldsymbol \alpha}^a$, $B_{\lambda {\cal C}}$ and $D_{\lambda {\cal C}}$, respectively.  Now
the fields are
\begin{eqnarray}
\label{colorfields2}
\Psi_{ {\cal C}} (\boldsymbol{k})   & = & {\cal U}_{\lambda} (\boldsymbol{k}) B_{\lambda {\cal C}}(\boldsymbol{k)}  + {\cal V}_{\lambda} (-\boldsymbol{k})
    D^\dag_{\lambda {\cal C}}(\boldsymbol{-k)}  \nonumber  \\
{\bf A}^a({\bs{x}}) &=&  \int
\frac{d{\bs{k}}}{(2\pi)^3}
\frac{1}{\sqrt{2\omega_k}}[{\boldsymbol \alpha}^a({\bs{k}}) + {\boldsymbol \alpha}^{a\dag}(-{\bs{k}})]
e^{i{\bs{k}}\cdot
{\bs {x}}}  \nonumber
\\
{\bf \Pi}^a({\bs{x}}) &=& \hspace{-.2cm}-i \int \!\!\frac{d{\bs{k}}}{(2\pi)^3}
\sqrt{\frac{\omega_k}{2}}
[{\boldsymbol \alpha}^a({\bs{k}})-{\boldsymbol \alpha}^{a\dag}(-{\bs{k}})]e^{i{\bs{k}}\cdot
{\bs{x}}}  \  , \nonumber
\end{eqnarray}
where $\omega_k = k e^{-2\Theta_k}$.
Note that the dressed quark expansion remains functionally
invariant with respect to the bare case since the quasi-particle
spinors have the inverse rotation
\begin{eqnarray} \label{eq:new spinors}
    {\cal U}_\lambda(\boldsymbol{k}) &=& \cos\frac{\theta_k}{2}
    u_{\lambda}(\boldsymbol{k}) - \lambda \sin\frac{\theta_k}{2}
    v_{\lambda}(-\boldsymbol{k}) \nonumber \\
     &= &\frac{1}{\sqrt{2}} \left[ \begin{array}{c}
        \sqrt{1+\sin \phi_k} \; \;  \chi_\lambda \\
        \sqrt{1-\sin \phi_k} \; \boldsymbol{\sigma} \cdot {\bf \hat{\bs k}} \;  \; \chi_\lambda\\
        \end{array} \right] \\ \nonumber
    {\cal V}_\lambda(-\boldsymbol{k}) &=& \cos\frac{\theta_k}{2}
    v_{\lambda}(-\boldsymbol{k}) + \lambda \sin\frac{\theta_k}{2}
    u_{\lambda}(\boldsymbol{k}) \\
    &= &\frac{1}{\sqrt{2}} \left[ \begin{array}{c}
         -\sqrt{1-\sin \phi_k} \; \boldsymbol{\sigma} \cdot { \bf \hat{\bs k}} \; \; \chi_\lambda\\
        \sqrt{1+\sin \phi_k} \; \; \chi_\lambda\\
        \end{array} \right].
\end{eqnarray}
Here  the quark gap angle, $\phi_k = \phi (k)$, is related to the
BCS angle $\theta_k$ by $tan(\phi_k  - \theta_k ) = m/k$.
The quasi-particle (BCS) vacuum, defined by $B_{\lambda {\cal C}} |\Omega
\rangle = D_{\lambda {\cal C}} |\Omega\rangle = \alpha^a_\mu |\Omega\rangle = 0$,  is
connected to the bare parton one, $b_{\lambda {\cal C}} |0\rangle = d_{\lambda {\cal C}} |0\rangle = a^a_\mu
|0\rangle = 0$, by
\begin{equation}
    |\Omega_{quark}\rangle =e^{- \int \!\!
    \frac{d\boldsymbol{k}}{(2\pi)^3}\lambda
    \tan\frac{\theta_k}{2} b^\dag_{\lambda {\cal C}}(\boldsymbol{k})
    d^\dag_{\lambda {\cal C}} (-\boldsymbol{k})} |0\rangle   \nonumber
\end{equation}
\begin{equation}
    |\Omega_{gluon}\rangle = e^ {- \!\! \int \!\!
    \frac{d\boldsymbol{k} }{(2\pi)^3} \frac{1}{2}\tanh\Theta_k D_{\mu \mu'}({\bs k})
    a_{\mu}^{a\dag} (\boldsymbol{k}) a_{\mu'}^{a\dag} (-\boldsymbol{k})} |0\rangle \ . \nonumber
\end{equation}
The BCS vacuum, $|\Omega \rangle = |\Omega_{quark}
\rangle \otimes |\Omega_{gluon} \rangle $, now contains
quark and gluon condensates (correlated $q\bar{q}$ and $gg$ Cooper pairs).
 Performing a variational minimization of the vacuum
expectation value of the Hamiltonian, $\delta \langle\Omega|H_{\rm eff}|
\Omega\rangle = 0$, independently with respect to $\phi_k$ and $\omega_k$,
yields the mass gap equations for each sector
\ba
    k s_k -  m c_k =&& \frac{2}{3}\int \!\!
    \frac{d\boldsymbol{q}}{(2\pi)^3} \left[
    ( s_k c_q x - s_q c_k)
V(|\boldsymbol{k-q}|)
    \right.  \nonumber \\  -&& \hspace{-.2cm} \left.
    2 c_k s_q {U}(\ar {\bs k} - {\bs q}\ar)  + 2c_q s_k {W}(\ar {\bs k} - {\bs
q}\ar) \right]
\ea
\ba
\label{ggapeq}
    \omega_k^2 &=& k^2-\frac{3}{4}\int \!\!
    \frac{d\boldsymbol{q}}{(2\pi)^3}
V(|\boldsymbol{k-q}|)[1+x^2]
    \bigg( \frac{\omega_q^2-\omega_k^2}{\omega_q} \bigg) \nonumber \\
    &+&  \frac{3}{4} \; g^2\int \!\!
    \frac{d\boldsymbol{q}}{(2\pi)^3} \frac{1 - x^2}{\omega_q} \ ,
\ea
where
 \begin{equation}
 {W}({|\bs k} - {\bs q}|) \equiv {U}(|{\bs k} - {\bs q}|)
\frac{x(k^2+q^2)-qk(1+x^2)}{\ar{\bs k}-{\bs q}\ar^2} \  ,
\end{equation}
with $s_k = sin \phi_k$, $c_k = cos \phi_k$ and $x = {\bs k
}\cdot {\bs q}$. The last term in Eq. (\ref{ggapeq}) originates
from the non-abelian component of the gluon kinetic energy.
Dimensional analysis of the above integrals reveals that the first
equation is UV finite for the linear potential since
$V_L(|\boldsymbol{p}|) = - 8 \pi \sigma /p^{4}$, but
not for the Coulomb potential
$V_C(|\boldsymbol{p}|) = - 4 \pi \alpha_s /p^{2}$. In
Eq. (\ref{ggapeq}) there are both logarithmical and quadratical
divergences in the UV region and an integration cutoff, $\Lambda
= 4$ GeV, determined in previous studies is used.

Once the current quark masses are fixed, the gap equations can be
solved numerically for the quark and   gluon
gap  angles.
Using $|q\rangle = B_{\lambda {\cal C}}(\boldsymbol{k})^\dag
|\Omega\rangle$ and $|g\rangle = \alpha^a_{\mu}(\boldsymbol{k})^\dag
|\Omega\rangle$, the quark and gluon self-energies are respectively


\ba
\label{eq:qself-energy}
  \epsilon_k & \equiv & \langle q | H_{\rm eff} | q \rangle = m s_k +k c_k - \nonumber \\
    \frac{2}{3} \hspace{-.2cm}&\int & \hspace{-.15cm} \!\!
    \frac{d\boldsymbol{q}}{(2\pi)^3} \left[
   [ s_k s_q + c_q
    c_k x] V(|\boldsymbol{k-q}|) \right. \nonumber  \\   &+& \left.
    \!\!\!\!2 s_k s_q {U}(\ar {\bs k} - {\bs q}\ar)  + 2c_q c_k {W}(\ar {\bs k} - {\bs
q}\ar) \right]
\ea

and, for fixed color index $a$  (no sum),
\ba
\varepsilon^{\mu \mu'}_k &\equiv & \langle \Omega|\alpha^a_\mu({\bs k}) \, H_{\rm eff} \,  \alpha^a_{\mu'}({\bs k})^\dagger |\Omega\rangle = \frac{ \omega^2_k +
k^2}{2\omega_k}  \delta_{\mu \mu'} \nonumber \\
&-&\frac{3}{4}\int \frac{d{\bs{q}}}{(2\pi)^3} V(\vert {\bs k}-{\bs q}
\vert) \frac{\omega_k^2+\omega_q^2}{\omega_q \omega_k} D_{\mu \mu'}({\bs
q}) \nonumber \\
&+&\frac{9}{2}g^2
\int \frac{d{\bs q}}{(2\pi)^3} \frac{1}{2\omega_k \omega_q}
\times
\nonumber \\
&&\left[
 2 D_{\mu \mu'}({\bs k}) -D_{\mu \nu} ({\bs k}) D_{\nu \nu' }({\bs q}) D_{\nu' \mu'}({\bs k})
\right] \ ,
\label{eq:gself-energy}
\ea
both of which are infrared divergent in the presence of an infrared
enhanced kernel. This is a welcomed feature of this
approach, as colored states are removed from the spectrum. The infrared
divergence cancels however in bound state equations for color singlet
states leading to a physical spectrum of mesons and baryons.

\section{Hybrid Mesons}

In previous publications~\cite{LC1,LC2,LCSS,LBC} we have used this
model to study the two-body meson and glueball systems by
diagonalizing $H_{\rm eff}$ using the Tamm-Dancoff and  Random
Phase approximations.  We also made predictions for three-body
glueballs (oddballs) \cite{LBC} and published \cite{LChybrid} a
brief study of the three-body hybrid  meson using a
variational treatment. We now extend the latter and also provide
more complete details of the variational calculation.

\subsection{Wavefunction ansatz and quantum numbers}

Following  our initial study \cite{LChybrid}, we work in the
hybrid center of momentum system and denote the momenta of the
dressed quark, anti-quark and gluon by ${\bs q}$, $ { \bar{\bs q}}$
and ${\boldsymbol g}$, respectively. We then define ${{\bs q}_+
\equiv  \frac{{\bs q}+\overline{\bs q}}{2}}$, ${\bs q_- \equiv q-
\overline{q} }$ and note that ${\boldsymbol g = -\bs q -
\overline{q} }=-2{\bs q}_+ $.

The color structure of a $q\bar{q}g$ hybrid is determined by
SU(3) algebra
\ba
    (3 \otimes \overline{3}) \otimes 8 &=& (8 \oplus 1) \otimes
    8 = ( 8\otimes 8) \oplus (8\otimes 1) \nonumber \\
    &=& 27 \oplus 10 \oplus 10 \oplus 8 \oplus 8
    \oplus 1 \oplus 8.
\ea
Note for an overall color singlet  the
quarks must  be in an octet state like the gluon.  As discussed
below, this leads to a repulsive $q \bar{q}$ interaction, confirmed by lattice at short range, which
raises the mass of the hybrid meson. The hybrid wavefunction will
therefore involve the color structure $T_{{\cal C}_1{\cal C}_2}^a
B^ \dag_{{\cal C}_1} D^ \dag_{{\cal C}_2} \alpha^{a \dag}$ and has
the general form
  \ba
    |\Psi^{JPC}\rangle = \int \!\! \int \!\! \frac{d\boldsymbol{q}_+}{(2\pi)^3}
    \frac{d\boldsymbol{q}_-}{(2\pi)^3} \Phi^{JPC}_{\lambda_1 \lambda_2 \mu}(\boldsymbol{q}_+,\boldsymbol{q}_-) \times \nonumber
    \\ T_{{\cal C}_1{\cal C}_2}^a B^ \dag_{\lambda_1{\cal C}_1}(\boldsymbol{q})
    D^ \dag_{\lambda_2{\cal C}_2}(\overline{\boldsymbol{q}}) \alpha^{a\dag}_{\mu}({ \boldsymbol g}) |\Omega \rangle \ ,
  \ea
which is summed over color and angular momentum magnetic sub-states.


There are five  angular momenta in this system,  two orbital,
${\bf l}_\pm$ (associated with ${\bs q}_\pm$) having $z$
projections $m_\pm$, and the 3 spins, ${S}_q = {
S}_{\overline{q}}=1/2$ and ${S}_g=1$ with projections $\lambda_1$,
$\lambda_2$ and $\mu$, respectively. To form states with total
angular momentum $J$, projection $m_J$, we use the  coupling
scheme,
  ${\boldsymbol   S} = \bs S_q + \bs S_{\overline{q}},
   \bs  j = \bs S_g + \bs l_+   ,
   \bs L  =\bs j +\bs  l_-  ,
   \bs  J = \bs L +{\boldsymbol  S }$.
Then with  the appropriate Clebsch-Gordan coefficients, the hybrid
wavefunction can be expressed in terms of a radial wavefunction
$F^{JPC}(q_+, q_-)$ and spherical harmonics, $Y_{l_{\pm}}^{m_{\pm}}(\bs q_{\pm})$,
\begin{eqnarray}
    \Phi^{JPC}_{\lambda_1 \lambda_2 \mu}(\boldsymbol{q}_+,\boldsymbol{q}_-)=
    F^{JPC}(q_+,q_-) Y_{l_+}^{m_+}(\bs  {\hat q}_+) Y_{l_-}^{m_-}({\bs {\hat q}}_-) \times && \nonumber \\
      (-1)^{\frac{1}{2}-\lambda_2}
     \langle \frac{1}{2} \frac{1}{2}, \lambda_1
    (-\lambda_2) | S m_S \rangle
    (-1)^{\mu} \langle 1 l_+, (-\mu) m_+ |
    jm_j \rangle \times&& \nonumber \\
  \langle j l_-, m_j m_- | Lm_L \rangle \langle L
    S, m_L m_S | J M_J \rangle  \  . \; \; \; \;  \; \; \; \; \;  && \nonumber
  \end{eqnarray}


\begin{table} [h]
\begin{ruledtabular}
  \begin{tabular}{|c|c|c|c|c|c|c|c|c|c|}
    $l_+$ & $l_-$ & $S$ & $j$ & $L$ & $J$ & $P$ & $C$ & $J^{PC}$ & \\
    \hline
    \hline
    0 & 0 & 0 & 1 & 1 & 1 & + & - & 1$^{+-}$ & \\
    0 & 0 & 1 & 1 & 1 & 0 & + & + & 0$^{++}$ & \\
    0 & 0 & 1 & 1 & 1 & 1 & + & + & 1$^{++}$ & \\
    0 & 0 & 1 & 1 & 1 & 2 & + & + & 2$^{++}$ & \\
    \hline
    0 & 1 & 0 & 1 & 0 & 0 & - & + & 0$^{-+}$ & \\
    0 & 1 & 0 & 1 & 1 & 1 & - & + & 1$^{-+}$ & Exotic \\
    0 & 1 & 0 & 1 & 2 & 2 & - & + & 2$^{-+}$ & \\
    0 & 1 & 1 & 1 & 0 & 1 & - & - & 1$^{--}$ & \\
    0 & 1 & 1 & 1 & 1 & 0 & - & - & 0$^{--}$ & Exotic \\
    0 & 1 & 1 & 1 & 1 & 1 & - & - & 1$^{--}$ & \\
    0 & 1 & 1 & 1 & 1 & 2 & - & - & 2$^{--}$ & \\
    0 & 1 & 1 & 1 & 2 & 1 & - & - & 1$^{--}$ & \\
    0 & 1 & 1 & 1 & 2 & 2 & - & - & 2$^{--}$ & \\
    0 & 1 & 1 & 1 & 2 & 3 & - & - & 3$^{--}$ & \\
    \hline
    1 & 0 & 0 & 0 & 0 & 0 & - & - & 0$^{--}$ & Forbidden \\
    1 & 0 & 0 & 1 & 1 & 1 & - & - & 1$^{--}$ & \\
    1 & 0 & 0 & 2 & 2 & 2 & - & - & 2$^{--}$ & \\
    1 & 0 & 1 & 0 & 0 & 1 & - & + & 1$^{-+}$ & Forbidden \\
    1 & 0 & 1 & 1 & 1 & 0 & - & + & 0$^{-+}$ & \\
    1 & 0 & 1 & 1 & 1 & 1 & - & + & 1$^{-+}$ & Exotic \\
    1 & 0 & 1 & 1 & 1 & 2 & - & + & 2$^{-+}$ & \\
    1 & 0 & 1 & 2 & 2 & 1 & - & + & 1$^{-+}$ & Exotic \\
    1 & 0 & 1 & 2 & 2 & 2 & - & + & 2$^{-+}$ & \\
    1 & 0 & 1 & 2 & 2 & 3 & - & + & 3$^{-+}$ & Exotic \\
  \end{tabular}
  \caption{Hybrid meson quantum numbers up to $J = 3$. Note  exotic states and  states
  forbidden by transversality: $l_+=1$ cannot couple to $j=0$.}
  \label{table:states}
\end{ruledtabular}
\end{table}


Since the intrinsic parity for a $q {\bar q}$ pair and a gluon are both
$-1$, and the two orbital parities are $(-1)^{l_-}$ and
$(-1)^{l_+}$, the total hybrid meson parity is
\begin{equation}
    P = (-1) (-1) (-1)^{l_+} (-1)^{l_-} = (-1)^{l_+ + l_-}.
\end{equation}

Finally, exchanging all additive quantum numbers, as required
by charge conjugation, yields a $(-1)^{l_-+S}$ factor
from the space and spinor $q\overline{q}$ components
which needs to be combined with the phase of the $q\overline{q}g$ composite
color component. Although the gluon octet is not an eigenstate of
C-parity,  each gluon has a
$q\overline{q}$ octet partner with opposite C-parity, resulting
in a $-1$ contribution for the combined $[[3 \otimes \bar{3}]_8\otimes
8]_1$ system. Therefore the hybrid C-parity is
\begin{equation}
    C = (-1) (-1)^{{l_-} +S} = (-1)^{1+l_-+S}.
\end{equation}
The extra $-1$ phase, as compared to a conventional $q\overline q$
meson having C-parity $(-1)^{l_-+S}$, is responsible for generating exotic quantum
numbers for certain hybrid  states (e.g. $J^{PC}=1^{-+}$).  Table \ref{table:states}
lists  quantum numbers for the model  hybrid states for $J$ up to 3.
Note the exotic quantum number states and also states forbidden by
the Coulomb gauge transversality condition (gluon orbital $l_+ =
1$ can not couple with its spin to produce $j = 0$).

\subsection{Variational equations of motion}

We now compute the hybrid mass, $M_{J^{PC}}$,  for each $J^{PC}$
with special interest focusing upon the exotic states. In terms
of the above variational wavefunction, and upper bound for the mass is given by
\ba
M_{J^{PC}}& =& \frac{ \langle \Psi^{JPC} | H_{\rm eff} | \Psi^{JPC} \rangle} {\langle\Psi^{JPC}|\Psi^{JPC}\rangle} \nonumber \\
&=&  {M}_{self} + M_{q\overline{q}} + M_{qg} + M_{qgq} +
  M_{ggg} \ .
\ea
Here, the subscripts indicate the mass contribution from  the
self-energy of the three constituents, $M_{self}$,  the
$q\overline{q}$ interaction, $M_{q\overline{q}}$, the $qg$ and
$\overline{q}g$ interactions, $M_{qg}$, the $2^{nd}$ order
correction from the $qgq$ and $\overline{q}g\overline{q}$
vertices,  $M_{qgq} $, and the $2^{nd}$ order correction from triple
gluon vertices, $M_{ggg} $. The three-body expectation value
entails twelve dimensional integrals which can be reduced to nine
dimensions by working in the center of momentum. The
detailed expressions are
\begin{widetext}
  \begin{eqnarray} \label{eq:self-energy}
  { M_{self} = \int\!\!\int\!\! \frac{d\boldsymbol{q}}{(2\pi)^3}
  \frac{d\overline{\boldsymbol{q}}}{(2\pi)^3}
  \Phi_{\lambda_1 \lambda_2 \mu}^{JPC \dag}(\boldsymbol{q},\overline{\boldsymbol{q}})
  \Phi_{\lambda_1 \lambda_2 \mu'}^{JPC}(\boldsymbol{q},\overline{\boldsymbol{q}}) }  \Bigg[D_{\nu \nu'}({\boldsymbol g})
  (\epsilon_{\boldsymbol{q}}+\epsilon_{\overline{\boldsymbol{q}}}) +
  D_{\mu \nu}({\boldsymbol g})
  D_{ \mu' \nu'}({\boldsymbol g})
  \varepsilon^{\nu \nu'}_{\boldsymbol g}\Bigg]
  \end{eqnarray}

  \begin{eqnarray}
  \lefteqn{ \hspace{-1cm} M_{q\overline{q}}=-\frac{1}{2}\int\!\!\int\!\!\int\!\! \frac{d\boldsymbol{q}}{(2\pi)^3}
  \frac{d\overline{\boldsymbol{q}}}{(2\pi)^3} \frac{d\boldsymbol{q'}}{(2\pi)^3}
  \Phi_{\lambda_1 \lambda_2 \mu}^{JPC \dag}(\boldsymbol{q},\overline{\boldsymbol{q}})
  \Phi_{\lambda_1' \lambda_2' \mu'}^{JPC}(\boldsymbol{q'},\boldsymbol{q}+\overline{\boldsymbol{q}}-\boldsymbol{q'})
  D_{\mu \mu'}(\boldsymbol{g}) } \nonumber\\
  && \Bigg[ \frac{1}{3}
V(|\boldsymbol{q'-q}|)
  {\cal U}_{\lambda_1 \boldsymbol{q}}^\dag {\cal U}_{\lambda_1' \boldsymbol{q'}}
  {\cal V}_{\lambda_2' \boldsymbol{q+\overline{q}-q'}}^\dag
  {\cal V}_{\lambda_2 \boldsymbol{\overline{q}}}+
V(|\boldsymbol{q}+\overline{\boldsymbol{q}}|)
  {\cal U}_{\lambda_1 \boldsymbol{q}}^\dag {\cal V}_{\lambda_2 \boldsymbol{\overline{q}}}
  {\cal V}_{\lambda_2' \boldsymbol{q+\overline{q}-q'}}^\dag
  {\cal U}_{\lambda_1' \boldsymbol{q'}}\Bigg]
  \end{eqnarray}

  \begin{eqnarray}
    {M}_{qg} &=& \frac{3}{4}\int\!\!\int\!\!\int\!\! \frac{d\boldsymbol{q}}{(2\pi)^3}
    \frac{d\overline{\boldsymbol{q}}}{(2\pi)^3}
    \frac{d\boldsymbol{q'}}{(2\pi)^3 } \nonumber \\
    && \Bigg[ \frac{\omega_{q+\overline{q}}+\omega_{q'+\overline{q}}}{\sqrt{\omega_{q+\overline{q}}
  \, \,  \omega_{q'+\overline{q}}}} \Phi_{\lambda_1 \lambda_2 \mu}^{JPC
    \dag}(\boldsymbol{q},\overline{\boldsymbol{q}})
    \Phi_{\lambda \lambda_2 \mu'}^{JPC}(\boldsymbol{q'},\overline{\boldsymbol{q}}) D_{\mu'
    \mu''}(\boldsymbol{q}+\overline{\boldsymbol{q}}) D_{\mu
    \mu''}(\boldsymbol{q}+\overline{\boldsymbol{q}})
V(|\boldsymbol{q-q'}|)
    {\cal U}_{\lambda \boldsymbol{q'}}^\dag {\cal U}_{\lambda_1 \boldsymbol{q}}
     \nonumber \\
    && + \frac{\omega_{q+\overline{q}}+\omega_{q+q'}}{\sqrt{\omega_{q+\overline{q}}
   \, \,  \omega_{q+q'}}}
    \Phi_{\lambda_1 \lambda_2 \mu}^{JPC
    \dag}(\boldsymbol{q},\overline{\boldsymbol{q}})
    \Phi_{\lambda_1 \lambda \mu'}^{JPC}(\boldsymbol{q},\boldsymbol{q'}) D_{\mu'
    \mu''}(\boldsymbol{q}+\boldsymbol{q'}) D_{\mu''\mu}(\boldsymbol{q}+
    \overline{\boldsymbol{q}})
    V(|\boldsymbol{\overline{q}}-\boldsymbol{q'}|) {\cal V}_{\lambda \boldsymbol{q'}}^\dag
    {\cal V}_{\lambda_2 \overline{\boldsymbol{q}}}
    \Bigg]
  \end{eqnarray}

  \begin{eqnarray}
    \lefteqn{
    \!\!\!\!\!\!\!\!\!\!\!\!\!\!\!\!\!\!\!\!\!\!\!\!\!\!\!\!\!\!\!\!
   \hspace{1.5cm} {M}_{qgq} = \frac{1}{2} \int\!\!\int\!\!\int\!\! \frac{d\boldsymbol{q}}{(2\pi)^3}
    \frac{d\overline{\boldsymbol{q}}}{(2\pi)^3}
    \frac{d\boldsymbol{q'}}{(2\pi)^3}
    \Phi_{\lambda_1 \lambda_2 \mu}^{JPC \dag} (\boldsymbol{q},\overline{\boldsymbol{q}})
    \Phi_{\lambda_1' \lambda_2' \mu'}^{JPC}(\boldsymbol{q'},\boldsymbol{q}+\overline{\boldsymbol{q}}-\boldsymbol{q'})
    D_{\mu \mu'}(\boldsymbol{g})} \nonumber \\
    && \Bigg[ \frac{1}{3}
    {U}_{m n}(\boldsymbol{q'}-\boldsymbol{q})
    {\cal U}_{\lambda_1 \boldsymbol{q}}^\dag \alpha_m {\cal U}_{\lambda_1' \boldsymbol{q'}}
    {\cal V}_{\lambda_2' \boldsymbol{q+\overline{q}-q'}}^\dag \alpha_n {\cal V}_{\lambda_2 \boldsymbol{\overline{q}}}
    +
    {U}_{m n}(\boldsymbol{q}+\overline{\boldsymbol{q}})
    {\cal U}_{\lambda_1 \boldsymbol{q}}^\dag \alpha_m {\cal V}_{\lambda_2 \boldsymbol{\overline{q}}}
    {\cal V}_{\lambda_2' \boldsymbol{q+\overline{q}-q'}}^\dag \alpha_n {\cal U}_{\lambda_1' \boldsymbol{q'}}
    \Bigg]
  \end{eqnarray}

  \begin{eqnarray}
   &&{{M}_{ggg} = \frac{3}{8}i \int\!\!\int\!\!\int\!\! \frac{d\boldsymbol{q}}{(2\pi)^3}
    \frac{d\overline{\boldsymbol{q}}}{(2\pi)^3} \frac{d\boldsymbol{q'}}{(2\pi)^3}
    \frac{1}{\sqrt{\omega_{-q'-\overline{q}} \, \omega_{q+\overline{q}}}}
    \Phi_{\lambda_1 \lambda_2 \mu}^{JPC \dag}(\boldsymbol{q},\overline{\boldsymbol{q}})
    \Phi_{\lambda_1' \lambda_2 \mu'}^{JPC}(\boldsymbol{q'},\overline{\boldsymbol{q}}) }
     ({\cal U}_{\lambda_1 \boldsymbol{q}}^\dag \alpha_h {\cal U}_{\lambda_1' \boldsymbol{q'}}+
    {\cal V}_{\lambda_1 \boldsymbol{q}}^\dag \alpha_h {\cal V}_{\lambda_1' \boldsymbol{q'}}) \nonumber \\
    && \hspace{2cm} \Bigg\{ \nabla_l {U}_{k h}(\boldsymbol{q}-\boldsymbol{q'})
    \Big[ D_{\mu k}(\boldsymbol{g})
    D_{l \mu'}(-\boldsymbol{q'}-\overline{\boldsymbol{q}}) - D_{\mu l}(\boldsymbol{g})
    D_{k \mu'}(-\boldsymbol{q'}-\overline{\boldsymbol{q}}) \Big]  \nonumber \\
    && \hspace{2cm}  +\; i(\boldsymbol{q}-\boldsymbol{q'})_l {U}_{l h}
    (\boldsymbol{q'} + \boldsymbol{q})
    D_{\mu k}(\boldsymbol{g})
    D_{k \mu'}(-\boldsymbol{q'}-\overline{\boldsymbol{q}})
    \\
    && \hspace{2cm} -\; i {U}_{k h}(\boldsymbol{q}-\boldsymbol{q'})\Big[(\boldsymbol{q'} +\overline{\boldsymbol{q}})_l
    D_{\mu l}(\boldsymbol{g})
    D_{k \mu'}(-\boldsymbol{q'}-\overline{\boldsymbol{q}})  +
    (\boldsymbol{q}+\overline{\boldsymbol{q}})_l D_{\mu k}
    (\boldsymbol{g})
    D_{l \mu'}(-\boldsymbol{q'}-\overline{\boldsymbol{q}}) \Big]
    \Bigg\} \nonumber \ .
  \end{eqnarray}
\end{widetext}

\begin{figure} [b]
    \hspace{-0.5cm}
    \includegraphics[width=.45\textwidth]{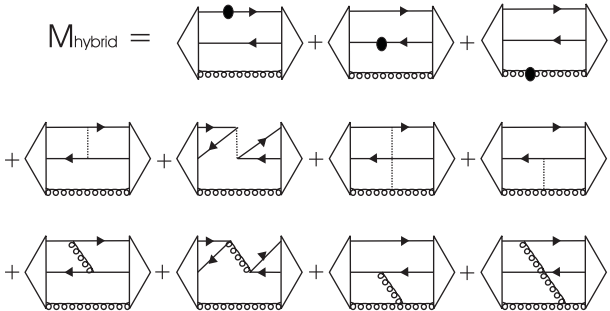}
    \caption{Diagrams for $\langle \Psi^{JPC} |H_{\rm eff}| \Psi^{JPC} \rangle$.}
    \label{fig:diagrams}
\end{figure}

In the above expressions, $\epsilon_{\boldsymbol{q}}$,
$\epsilon_{{\boldsymbol{\overline q}}}$ and
$\varepsilon^{\mu \mu'}_{\boldsymbol
{g}}$ are the quark, anti-quark and gluon self-energies,
respectively, evaluated at the indicated momentum (${\boldsymbol
{g} = -\boldsymbol{q}- {\boldsymbol{\overline q}}} $).  A
pictorial representation for each type of contribution is given by
the Feynman diagrams in Fig. \ref{fig:diagrams}.

The above expectation values are then computed variationally using
the separable  radial  wavefunction, $F(q_+,q_-)=f(q_+,\alpha_+)
f(q_-,\alpha_-)$, having two variational parameters, $\alpha_+$
and $\alpha_-$. We investigated two functional forms for $f$; a
gaussian and a scalable, numerical solution from our two body
meson studies. In general, the gaussian radial wavefunction,
\begin{equation}
f(q_{\pm}, \alpha_{\pm}) =e^{-x^2_\pm}, \; \; x_\pm = \frac{q_\pm}{\alpha_\pm}
\end{equation}
provided better results (lower variational mass) for s-wave states
when compared to the numerical one. This was also true for p-wave orbital excitations, provided the
gaussian was multiplied by $x_\pm$ corresponding to $l_\pm = 1$. All integrals were calculated
using the Monte Carlo method with the adaptive sampling algorithm
VEGAS \cite{Vegas}. The integrals were evaluated several times
with an increasing number of points until a weight-averaged result
converged. The hybrid mass error introduced by this procedure  is
about $\pm $50 MeV. For each $J^{PC}$ hybrid state we optimized
the variational parameters $\alpha_+$ and $\alpha_-$ to produce
the lowest mass. In terms of the string tension, their  values fell  in the ranges $0.9 \sqrt{\sigma} \leq
\alpha_+ \leq 1.7 \sqrt{\sigma}$ and $1.0 \sqrt{\sigma} \leq \alpha_- \leq 3.3 \sqrt{\sigma}$.

\section{Results: Hybrid Meson Spectrum}

\subsection{Light hybrid mesons}

For the light hybrid calculation we used $m = $ 5 MeV  \cite{PDG}
for  the $u/d$ current quark mass. Results  are listed in Table
\ref{table:spectrum-u} which shows the ground state is the $0^{++}$
non-exotic scalar, followed by the triplet
$1^{+-}$, $2^{++}$ and $1^{++}$. The lightest hybrid mass is
2.1 GeV.

For exotic states, as can be seen from Table \ref{table:states},
at least one p-wave in ${\bs q}_+$ or ${\bs q}_-$ is required.
Because the $q\bar{q}$ interaction is repulsive for quarks in a
color octet state, the excitation energy is less for a $l_+$
(gluon orbital) excitation than a $l_-$ ($q\bar{q}$ orbital)
excitation since the quarks are further separated and experience a
larger repulsive linear force.

\begin{figure} [b]
\includegraphics[width=0.45\textwidth]{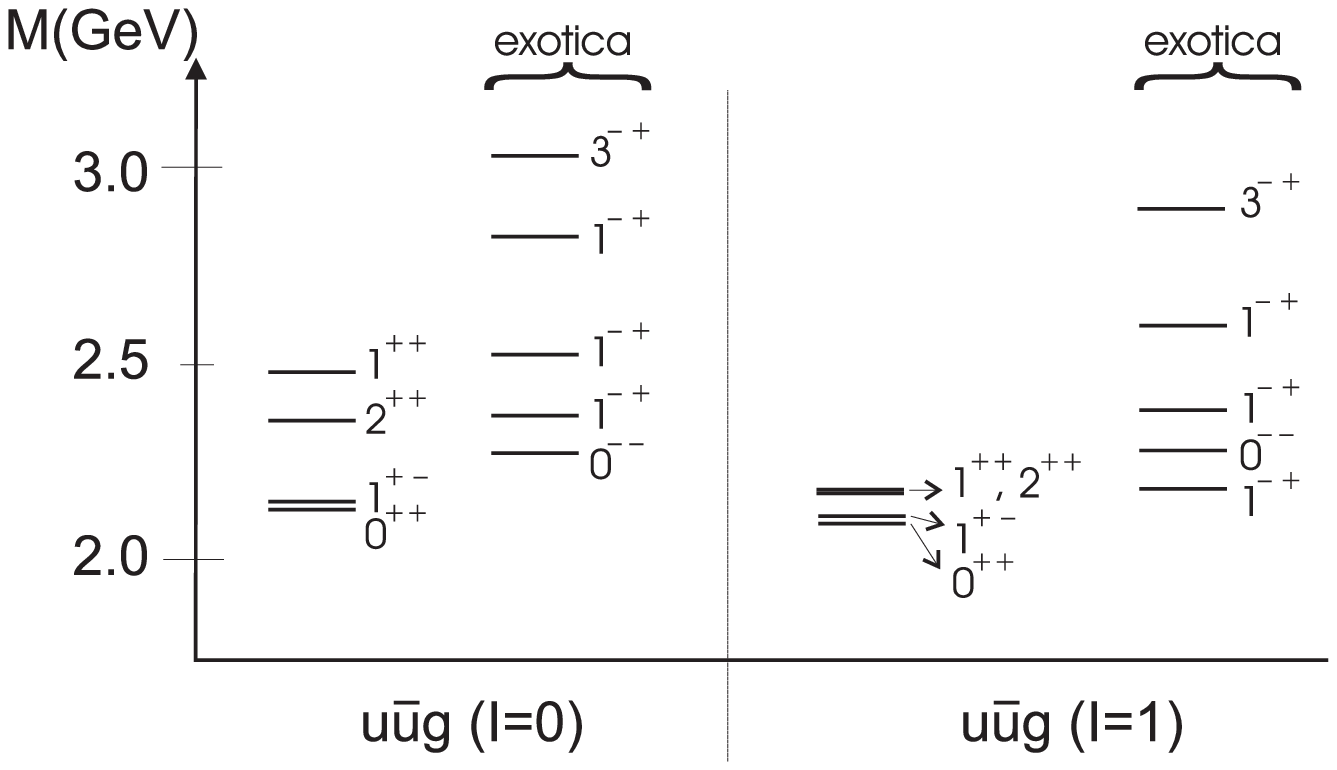}
  \caption{$u \overline{u}g$ spectrum.}
  \label{fig:resultsUU}
\end{figure}

\begin{table}
\begin{ruledtabular}
  \begin{tabular}{|c|c|c|l|}
    $(I)J^{PC}$ & $M_{J^{PC}}$ (MeV) & $M_{J^{PC}}$ (MeV) & \\
    & no corrections & with $g^2$ corrections & \\
    \hline
     \hline
    $(0)0^{++}$ & 2080 & 2135 & \\
    $(1)0^{++}$ & 2065 & 2100 & Ground \\

    \hline
    $(0)1^{+-}$ & 2135 & 2140 & $^*$ \\
    $(1)1^{+-}$ & 2135 & 2140 &  $^*$   \\
    \hline
    $(0)2^{++}$ & 2340 & 2335 & \\
    $(1)2^{++}$ & 2180 & 2170 & \\

       \hline
    $(0)1^{++}$ & 2415 & 2470 & \\
    $(1)1^{++}$ & 2110 & 2170 & \\
        \hline
         \hline
    $(0)1^{-+}$ & 2500 & 2525 & Exotic\\
    $(1)1^{-+}$ & 2205 & 2220 & Exotic\\
     \hline
    $(0)0^{--}$ & 2275 & 2280 & Exotic\\
    $(1)0^{--}$ & 2280 & 2285 & Exotic\\
    \hline
    $(0)1^{-+}$ & 2370 & 2400 & Exotic $^*$ \\
    $(1)1^{-+}$ & 2370 & 2400 & Exotic $^*$ \\
          \hline
    $(0)1^{-+}$ & 2760 & 2790 & Exotic\\
    $(1)1^{-+}$ & 2570 & 2600 & Exotic\\
    \hline
    $(0)3^{-+}$ & 3030 & 3040 & Exotic\\
    $(1)3^{-+}$ & 2910 & 2915 & Exotic\\
  \end{tabular}
  \caption{Spectrum of light hybrid meson states. Error $\approx \pm$ 50 MeV.  $^*$Isospin degenerate states.}
  \label{table:spectrum-u}
\end{ruledtabular}
\end{table}

The lightest exotic state is the $I=1,1^{-+}$, with mass 2.22 GeV.
This is slightly higher than  the Flux Tube model and lattice QCD
predicted masses for this state which were between 1.7 and 2.1 GeV
(see Table \ref{table:comparison-u}).

We studied the effects from including the non-abelian (NA) and
hyperfine  corrections for several states. Generally,  both effects were
small (except the hyperfine correction for charmed quarks, see below),  roughly of the same order as the overall 50 MeV Monte Carlo error.  In particular, the NA correction entailed several terms with different
signs which tended to cancel.

Our model exotic spectrum (see Fig. \ref{fig:resultsUU}) spans
almost a GeV,  between 2.1 and about 3 GeV, and includes
predictions for $J$ up to 3.  There are no
exotic $J =$  2 model states
in this region since they require a d-wave or two p-waves, both involving much higher excitations.

Finally, we comment on an interesting isospin splitting effect.
From Fig. \ref{fig:diagrams}, annihilation
terms only contribute to the hybrid mass if the $q \overline{q}$ pair
has quantum numbers consistent with the interaction. This is
satisfied when $I_{q\overline{q}}=0$ and $S=1$. The annihilation
diagrams can increase the $I = 0$ hybrid states by several hundred
MeV, as detailed in Table \ref{table:spectrum-u}.
In other cases,
the $1^{+-}$ and one of the $1^{-+}$ states should be isospin
degenerate, as we compute to within the Monte Carlo error. On the
other hand, the states $0^{++}$ and $0^{--}$ are not expected to
be degenerate but, within the error, they are. It may be that the
isospin splitting is not zero, but rather is smaller than the
numerical error. Note the annihilation process for s-wave,
isoscalar quarks in a triplet spin state is analogous to $e^- e^+$
annihilation in the triplet state of positronium.

\subsection{Strange hybrid mesons}

Table \ref{table:spectrum-s} summarizes results obtained for the
$s \bar{s}g$ (hidden strangeness) hybrid mesons using  a bare strange quark mass of 80 MeV.
Now, the ground state is given by the non-exotic pseudovector
state $1^{+-}$,  with a  2.125 GeV mass,
not at all reflecting the  75 MeV additional quark flavor mass
contribution (the hybrid calculation is only sensitive to current quark masses above 200 MeV).  Our prediction is in  good agreement with the
Flux Tube model and slightly above the only lattice prediction
(see Table \ref{table:comparison-u}). In the exotic sector, the
lightest state is given by $0^{--}$, with  mass 2.3 GeV.
Although there are also hybrid states with explicit strangeness, e.g.  $s\bar {u}g$,
we do not show predictions since the effect from the $s/u$ quark mass difference is small.

\begin{table}
\begin{ruledtabular}
  \begin{tabular}{|c|c|c|l|}
    $J^{PC}$ &$M_{J^{PC}}$ (MeV) & $M_{J^{PC}}$ (MeV) & \\
    & no corrections & with $g^2$ corrections & \\
    \hline
    \hline
    $1^{+-}$ & 2095 & 2125 & Ground \quad  \\
    \hline
    $0^{++}$ & 2045 & 2140 & \\
    \hline
    $2^{++}$ & 2290 & 2315 & \\
     \hline
    $1^{++}$ & 2325 & 2420 & \\
    \hline
    \hline
    $1^{-+}$ & 2350 & 2395 & Exotic\\
    \hline
    $0^{--}$ & 2270 & 2300 & Exotic\\
    \hline
    $1^{-+}$ & 2440 & 2485 & Exotic\\
    \hline
    $1^{-+}$ & 2760 & 2820 & Exotic\\
    \hline
    $3^{-+}$ & 2995 & 3030 & Exotic\\
  \end{tabular}
  \caption{Spectrum of $s\bar {s} g$ states. Error $\approx \pm $ 50 MeV.}
  \label{table:spectrum-s}
\end{ruledtabular}
\end{table}

\begin{table}
\begin{ruledtabular}
  \begin{tabular}{|c|c|c|l|}
   $J^{PC}$ & $M_{J^{PC}}$ (MeV) & $M_{J^{PC}}$ (MeV) & \\
    & no corrections & with $g^2$ corrections & \\
    \hline
    \hline
    $1^{+-}$ & 3310 & 3830 & Ground \quad  \\
    \hline
    $0^{++}$ & 3295 & 3945 & \\
     \hline
    $2^{++}$ & 3410 & 3965 & \\
    \hline
    $1^{++}$ & 3450 & 4100 & \\

    \hline
    \hline
    $1^{-+}$ & 3545 & 4020 & Exotic\\
    \hline
    $0^{--}$ & 3510 & 4020 & Exotic\\
    \hline
    $1^{-+}$ & 3590 & 4155 & Exotic\\
    \hline
    $1^{-+}$ & 3985 & 4565 & Exotic\\
    \hline
    $3^{-+}$ & 4065 & 4615 & Exotic\\
  \end{tabular}
  \caption{Spectrum of $c \bar{c}g$ states. Error $\approx \pm$ 50  MeV.}
  \label{table:spectrum-c}
\end{ruledtabular}
\end{table}

\subsection{Heavy hybrid mesons}

Table \ref{table:spectrum-c} shows the results for the $c \bar{c} g$ (charmonium)
hybrid mesons using a charmed quark mass of 1.0 GeV.
The ground state is given, again, by the $1^{+-}$ state, with
mass  3.83 GeV, while the lightest exotic hybrid lies
at 4.02 GeV. These numbers are in  reasonable agreement with
previous  lattice and Flux Tube predictions, as listed in Table
\ref{table:comparison-u}.

\begin{figure} [b]
   \includegraphics[width=.45\textwidth]{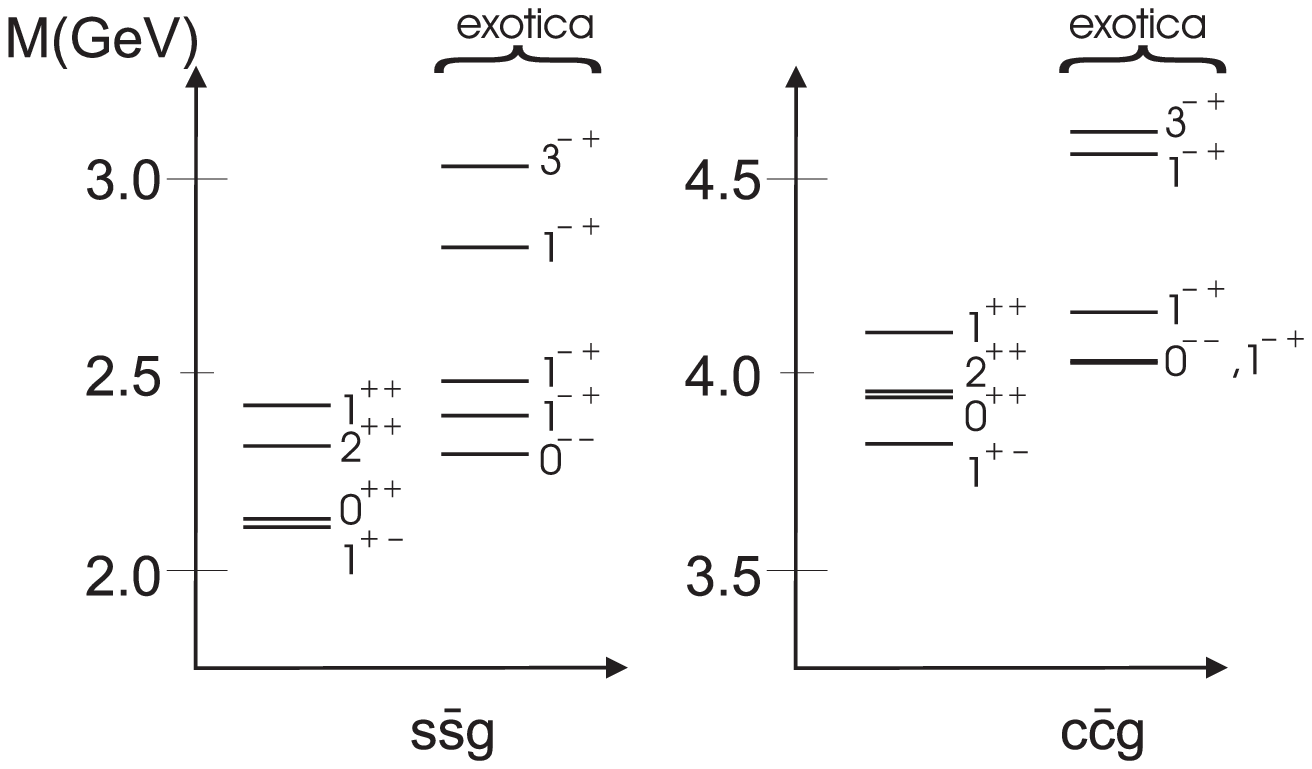}
   \caption{Low lying $s \overline{s}g$ and $c \overline{c}g$ spectra.}
   \label{fig:resultsSS-CC}
\end{figure}

Note that the correction introduced in the charmed case by the
$g^2$ terms is roughly 500 to 600 MeV, significantly higher than in the
lighter hybrid systems where the average corrections are 25 to 50
MeV. This large effect arises from the hyperfine correction to
the quark and anti-quark self-energies (see Eqs.
(\ref{eq:qself-energy}, \ref{eq:self-energy})), which is enhanced for
heavier quark masses as discussed further in  Ref. \cite{LCSS}. Related, and as illustrated in Fig.
\ref{fig:resultsSS-CC}, the charmonium hybrid spectrum now has a slightly
different level ordering from the lighter hybrid spectra.

\subsection{Sensitivity to potential and parameters}

One of our key findings using the Cornell potential is that  the mass of the lightest hybrid,
especially the exotic $1^{-+}$, is above 2 GeV.
Because of the ramifications of this result for
exotic state searches,
we have performed an interaction sensitivity study by varying both
 potential forms and parameters.

We first varied the parameters in the Cornell potential
to obtain a lower bound for our
predicted exotic hybrid mass.  Results are shown in Table \ref{table:varypot}
for different  Coulomb potential parameters, $\alpha_s$,
and string tensions, $\sigma$, found in the literature.   For
any combination of values
consistent with previous studies \cite{LC1,LC2,LCSS,LBC} it was not possible to
reduce the light hybrid mass to 1600 MeV. In
particular, we tried $0.0 \leq \alpha_s \leq 0.4$ and 367 MeV $
\leq \sqrt{\sigma} \leq 424$ MeV. Indeed, to obtain a hybrid mass
as low as 1600 MeV required an unphysical $\sqrt{\sigma}= $262
MeV.

\begin{table} [h]
\begin{ruledtabular}
  \begin{tabular}{|c|c|c|}
potential/parameters & $I = 1$   $u \bar{d} g$ hybrid &  $c \bar{c} g$ hybrid   \\
    \hline
    \hline
    Cornell [Eq. (\ref{2})] & &  \\
$\sqrt{ \sigma} = 367 \,  {\rm MeV}, \; \alpha_s = 0.4$ & 2220 &  4155 \\
$\sqrt{ \sigma} = 367 \,  {\rm MeV}, \; \alpha_s = 0.2$ & 2390 &  4415 \\
$\sqrt{ \sigma} = 367 \,  {\rm MeV}, \; \alpha_s = 0.0$ & 2540 &  4645 \\
$\sqrt{ \sigma} = 424 \,  {\rm MeV}, \; \alpha_s = 0.4$ & 2555 &  4525 \\
 \hline
Renormalized [Eq. (\ref{adampot})] & &  \\
$m_g = 526 \, {\rm MeV}$   & 2705 & 4730 \\
$m_g = 607 \, {\rm MeV}$   & 3010 & 5130 \\
  \end{tabular}
  \caption{Calculated light and charmonium hybrid  $1^{-+}$   masses, in GeV, using  different interactions.}
  \label{table:varypot}
\end{ruledtabular}
\end{table}

Table \ref{table:varypot} also lists predictions for the confining potential
given by Eq. (\ref {adampot}) for  values of the parameter $m_g = \sqrt{8 \pi \sigma/12.25}$
corresponding to the two different Cornell string tensions $\sigma$ but with the same current quark masses
($m_u =$ 5 MeV, $m_c =$ 1 GeV).
Note that this interaction yields $u \bar{d} g$ and $c \bar{c} g$ hybrids that are heavier  than those given by the Cornell potential. Most significantly, this potential also predicts the lightest exotic hybrid has
mass above 2 GeV.
If we use $m_c =$ 0.85 GeV, which provides a reasonable description of the charmonium spectrum, the $1^{-+}$ $c \bar{c}g$ mass decreases to 4815 MeV for  $m_g =$ 607 MeV.

\section{Searching for hybrid mesons}

Discovering exotic hadrons is a major goal motivating
the Jefferson Lab 12 GeV upgrade and is also being actively  pursued by
other collaborations and facilities, such as Babar, Belle, RHIC, etc.
For low energy investigations of light  quark exotic systems there is, unfortunately, no clean
energy scale demarkation, since
$\Lambda_{QCD}$ governs  the momentum distributions in light
mesons and the strange quark mass is of the same order of
magnitude.
The obvious detection strategy is therefore to perform statistically accurate cross sections
measurements
to  extract  partial wave amplitudes with explicitly exotic quantum numbers not
accessible to ordinary $q\bar{q}$ states.
However for (hidden) exotics with conventional meson quantum numbers, it will be
difficult to establish their nature.
Note certain  Flux Tube model \cite{Barnes} and lattice
predictions indicate that p-wave hybrid mesons prefer to decay
to hadron pairs with one hadron
also having a p-wave, rather than to two s-wave hadrons with a relative motion p-wave,
e.g. $\eta h_1$ in an s-wave as opposed to $\pi
\pi$ in a p-wave. It will be interesting to check this prediction
experimentally.

More germane to the results of  this paper are high energy
experiments where  novel tests  can be conducted based on
the scale separation provided by either the high beam  energy
or the heavy quark mass.  This is discussed in the
next two subsections.

\subsection{Application of dimensional counting rules}

Dimensional counting rules \cite{Brodsky} predict a power-law
production cross section behavior for a given state at
asymptotically high energies. They are based on the requirement
that in forming a bound state with an energetic quark, the other
partons must acquire very small relative momenta consistent
with the production hadron's internal momentum distribution.
This becomes highly unlikely in energetic collisions and therefore the production cross
section falls as a power law, with the exponent increasing with increasing
minimum number of constituents. For a hidden hybrid,
the wavefunction has  the Fock space expansion \be \ar \psi \ra =
\beta_1 \ar {c\bar{c}} \ra + \beta_2\ar {c\bar{c}g}  \ra + \cdot
\cdot  \cdot \ee where the quantum numbers are conventional but the
first coefficient $\beta_1$ is presumably small. Therefore the
power-law behavior will reveal the second term in this series and permit
distinguishing this exotic meson from ordinary $c\bar{c}$ charmonium (same
for  bottomonium). To be specific in the following discussion we
focus on the recently discovered $\psi(4260)$  whose nature
is currently under debate.  The same remarks apply  to any other
hidden  hybrid meson candidate.
\paragraph{Inclusive production}
First consider the inclusive production  reaction
$
e^- e^+ \to \psi + X $.
The virtual photon fragments into a $c\bar{c}$ pair, each carrying
half of the total center of mass momentum and
therefore each has an energy equal to the beam energy $E_{beam}$ (in a symmetric
collider).  The dimensional counting rules apply in the limit in
which the produced particle's energy approaches its threshold value,
that is, \be x= \frac{E_\psi}{E_{beam}}\to 1 \ . \ee In this
limit, the power law behavior for a conventional charmonium state
is $1-x$ and for a hybrid state with a minimum of one more gluon
constituent in the leading wavefunction, $(1-x)^3$
\cite{Gunionplb79}. That is, as the $\psi$ is produced with more
energy (this can be determined kinematically), the production cross
section decreases linearly towards the kinematical endpoint where
the $\psi$ has the maximum available energy.  Note that the hybrid
production cross section decreases even more
rapidly as a cubic polynomial.  With a sufficient number of events this
can be documented experimentally.

 \paragraph{Exclusive production}

Bodwin, Braaten and Lee \cite{Bodwinprl} have recently examined the
reaction
$
e^- e^+ \to \psi \psi
$.
Their study focused on the ground state $J/\psi$, but  their
arguments also apply to the production of excited
vector charmonia or a $J/\psi$ accompanied by a $\psi(4260)$. They
studied  $ \psi \psi $ production as a function of the (small) variable
( $m_c$ is the charmed quark mass)
$$
r=\frac{m_c}{E_{beam}}
$$
and find  in the limit $r\to 0$, the
differential cross section  at fixed angle is constant. However, a
straight-forward counting rule application  predicts the
production cross section is suppressed by a power of $s$ or,
equivalently, two powers of $r$, if one of the two produced hadrons
is predominantly a hybrid meson, that is \ba
\frac{d\sigma}{d\cos \theta} &\simeq& {\rm constant} \ , \; \; {\rm for} \; c\bar{c} \\
\frac{d\sigma}{d\cos \theta} &\simeq& \frac{\rm constant}{r^2} \ , \; \; {\rm for} \;
c\bar{c} g \ . \ea Establishing this behavior only requires
measuring double charmonium production  at
three sufficiently high energies. For non-vector mesons
both cross sections are further suppressed by additional powers of $r$,
but the extra $r^2$ signature  always marks the presence of
one more constituent, and therefore tags the hybrid meson. This
argument also applies to the production of light hybrids in
high-energy electron-positron colliders.

To conclude this subsection, we note that the power-law predictions
can be modified by QCD logarithms at very high energies and that in the strict limit
$r\to 0$ or $x\to 0$, any $c \bar c$  admixture in the
wavefunction would dominate  production.
However for  current, available low energies, the production
will still be dominated by the hybrid component if
$\beta_1<< \beta_2$, i.e.  if a predominantly pure hybrid state
is found.

\subsection{Distinguishing the hybrid from   charmonium}

In this subsection we propose a novel method applicable to the
decay of heavy quarkonium that enables identifying a
new state as either a radial excitation of conventional charmonium or a
hybrid state.  The method is based upon two key points.

\begin{figure} [t]
\psfig{figure=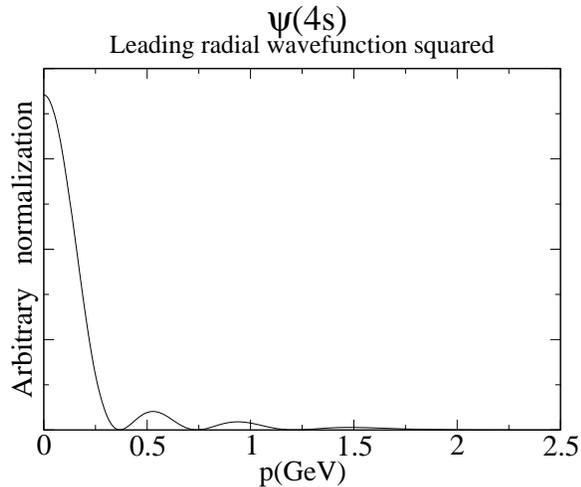,height=3.5in,angle=-90}
\caption{\label{4spdist} Probability density, $\ar \psi(p)\ar^2$, for the
4s charmonium state. This is the relative $c \bar{c}$ quark momentum  distribution.}
\end{figure}

\begin{figure} [t]
\psfig{figure=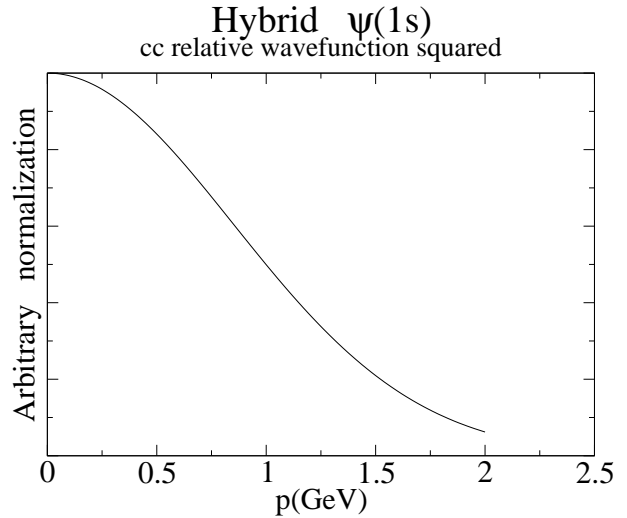,height=3.7in,angle=-90}
\caption{\label{1swave} Typical quark relative momentum
distribution in a $c \bar{c} g$ hybrid state with a mass near 4300
MeV. The extra mass  energy, relative to a $c \bar c$ $J/\psi$,
corresponds to gluon field excitations in collective models such
as the Flux Tube approach or in the quasi-particle (gluon) mass
gap in the constituent picture, but not in nodal radial excitation
for the relative $c\bar{c}$ motion (compare to the radially
excited charmonium distribution in  Fig. \ref{4spdist}).}
\end{figure}

First, we note that, due to the gluon mass  gap scale, a
conventional $q \bar{q}$ ground state  (e.g. a well established
$J/\psi$ or $\Upsilon$, etc.) is lighter than a $q \bar{q} g$
ground state hybrid with the same flavor. Indeed, the ground state
hybrid mass is more comparable to a radially excited quarkonium
state.  For example in our approach the $\psi(4s)$ and the ground
state vector $c\bar{c}g$ state have similar masses. Now different
eigenstates of a hermitian Hamiltonian are orthogonal with the
$n$th radial excited state having $n -1$ nodes. Therefore, even
though the total energies (masses) are similar, the relative
momentum distribution of the quarks in excited charmonium looks
quite different from the quark momentum distribution in the ground
state hybrid (see Figs. \ref{4spdist} and \ref{1swave}).

The second point involves the
Franck-Condon (FC) principle widely used in molecular physics.  Franck and Condon were
the first to appreciate that molecular
electronic transitions proceed too rapidly for the much heavier nuclei
to respond.   The FC principle is
applicable whenever there is a mass scale separation between
different particles. In the context of quarkonium this means that
the light fields (pions, gluons, etc.) quickly rearrange and
the heavy quarks do not appreciably change their momentum
distribution in the decay. Hence, the relative momentum
between the decay products directly correlates with the quark momentum distribution
in the parent quarkonium.
Unfortunately in the simplest 2-body decays such as
$$
\psi(ns)\to D\bar{D},\ \ \Upsilon(ns)\to B\bar{B}
$$
the FC constraint is not relevant since in the center of mass frame the momentum of the final
products is fixed. This leads to  smaller wavefunction overlaps
suppressing the decay somewhat.
However  in 3-body decays  such as
$$
\psi(ns)\to D\bar{D}\pi,\ \ \Upsilon(ns)\to B\bar{B}\pi\ ,
$$
the FC constraint applies.
The first reaction can be employed to study the recently discovered
$\psi(4260)$. The second will be useful in an envisioned
Belle collaboration measurement to
establish whether this excited bottomonium state  is the
predicted quark model 5s meson state.

\begin{figure} [h]
\psfig{figure=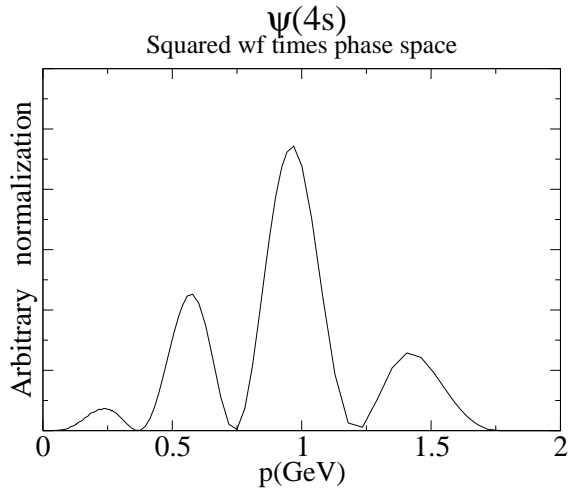,height=3.5in,angle=-90}
\caption{\label{4sphasedist} Momentum distribution  of Fig.
\ref{4spdist} multiplied by the phase space for the decay
$\psi(4260)\to D\bar{D} \pi$. This is the probability density for
finding   a $D\bar{D}\pi$ state with relative $D\bar{D}$ momentum
$p$ according to the Franck-Condon principle in the heavy-quark
limit. This signature will  be more robust for the related
bottomonium process $\Upsilon(5s)\to B\bar{B}\pi$.
 }
\end{figure}

Thus, we contend  that the relative momentum distribution between
the $D$ and $\bar{D}$ mesons in the  $D\bar{D}\pi$ system mirrors
the momentum distribution of the  quarks in the parent $\psi$
meson. Since the hybrid ground state wavefunction does not have a
node, the resulting momentum distribution for the $D\bar{D}$
subsystem is also node-less and thus smoother than that for a
conventional radially excited charmonium. Multiplying by the
relevant phase space distribution for this decay, yields the
momentum distribution in Fig. \ref{4sphasedist} that can be
observed experimentally in the heavy-quark limit. The maximum is
in the mid-momentum region where phase-space is larger and the
wavefunction is near a local maximum.

\begin{figure} [h]
\psfig{figure=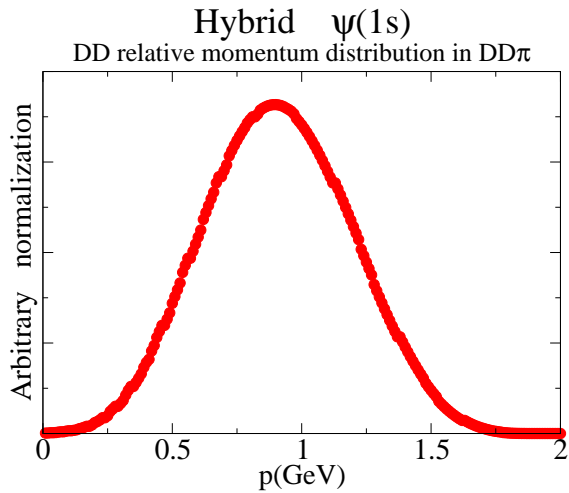,height=3.5in,angle=-90}
\caption{\label{1scomplete} The relative momentum distribution of
the $D\bar{D}$ pair in the $D\bar{D}\pi$ final state for a charmed
hybrid meson with momentum distribution given in Fig.
\ref{1swave}. The distribution of the final products has a smooth
bell shape in sharp contrast to the  radially excited quarkonium
distribution in Fig. \ref{4scomplete}.}
\end{figure}

\begin{figure}
\psfig{figure=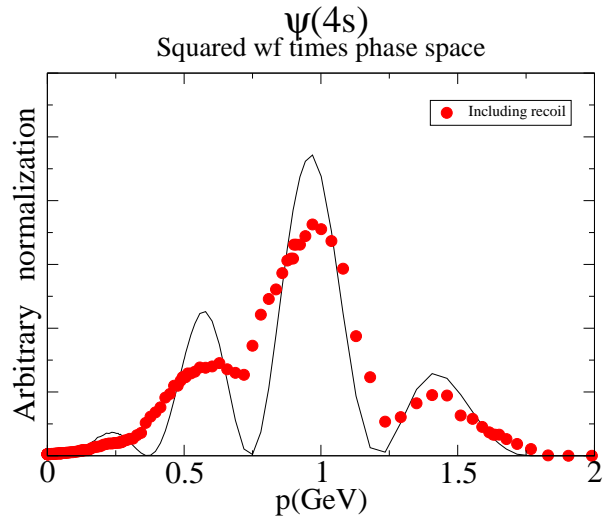,height=3.7in,angle=-90}
\caption{\label{4scomplete} The momentum distribution in Fig.
\ref{4sphasedist} (solid line) and the distribution (dots)
obtained averaging over a  150 MeV spread to approximate the
actual  quark momentum distribution in the $D$ meson. The
experimental signal, firmer for bottomonium than charmonium, is a
central peak with two shoulders for the relative momentum
distribution of the $B\bar{B}$ (or $D\bar{D}$) pair in the
$B\bar{B}\pi$ (or $D\bar{D}\pi$) final state. These adjacent
enhancements identify the resonance as a radially excited
quarkonium state as opposed to a hybrid meson.}
\end{figure}

However, since quarks are not infinitely heavy, the FC signature
is modified due to the recoil of the quarks in the $D$ meson,
yielding a different momentum distribution. For example, taking a
quark relative momentum between 150 and 200 MeV, one obtains the
$c \bar{c} g$ momentum distribution illustrated in Fig.
\ref{1scomplete} for the ground state vector hybrid, and the
smeared final state $D\bar{D}$ momentum distribution plotted in
Fig. \ref{4scomplete} for radially excited charmonium. As can be
seen, even after smearing, there is still residual structure
information adjacent to the central peak for radially excited
charmonium that is reminiscent of its parent charmonium
wavefunction behavior, in sharp contrast to the smooth,
bell-shaped hybrid distribution.

We therefore  advocate analyzing the $D\bar{D}$ and $B\bar{B}$
relative momentum distributions in $D\bar{D}\pi$ and $B\bar{B}\pi$
decays of highly excited quarkonia. Additional final state pions
or other light particles do not alter our arguments (but restrict
somewhat the available phase space), so there are several other
possible final state channels  to search.

\section{Discussion and Conclusions}

Our key model predictions are that the lightest  hybrid  mass is
2.1 GeV, with the lightest $1^{-+}$ exotic  state slightly above
2.2 GeV. Lattice and Flux Tube calculations yield a mass of at
least 1.8 GeV for the $1^{-+}$, except Ref.  \cite{Hedditch} which
predicts a mass around 1.7 GeV. Thus, with exception to this last
work, the composite model analyses appear to preclude the
possibility of the reported $1^{-+}$ exotica, $\pi_1(1400)$ and
$\pi_1(1600)$, being hybrid mesons. If this is correct, one should
investigate other structures for those two hadrons, such as
tetraquark molecules, with both $q\overline{q}$ pairs in color
singlets, or tetraquark atoms, where quark pairs are in
intermediate non-singlet color states and we are currently
applying our model to these systems. However,  if the lattice 1.7
GeV $1^{-+}$ prediction is robust, one can not yet exclude  the
observed $\pi_1(1600)$ from being a hybrid, but this  still does
not explain the structure of the $\pi_1(1400)$. It would be very
useful to have other lattice measurements, using the same
techniques as Ref. \cite{Hedditch}, to confirm or reject this
result. Related,  we have also varied our model parameters and
potential forms to obtain a  lower bound for our predicted exotic
hybrid mass which is clearly above 2 GeV.

Regarding isospin splitting, our results show an enhanced
splitting from $g^2$ corrections.  For the $0^{++}$ hybrid, the
corrections increased the splitting from 15 to 35 MeV,  the
maximum increase in the light hybrid spectrum.

In the strange sector, we predict the lightest non-exotic hybrid
mass is 2.125 GeV, while the lightest exotic mass is 2.30 GeV.
These values compare reasonably well with Flux Tube
\cite{Barnes,Close,Katja} and, slightly lighter, lattice
\cite{Hedditch} results. For  the charmed sector, our predictions
of 3.83 GeV for the lightest hybrid and 4.02 GeV for the lowest
exotic are also in good agreement with several other   lattice and
Flux Tube studies (see Table \ref{table:comparison-u}).

As mentioned above, the different $g^2$ corrections produced an
overall small effect, about the same order as the Monte Carlo
error. However, the hyperfine correction becomes  important for
heavier quark mass.  In the charmed sector, this correction added
about 500 to 600 MeV to the hybrid mass. Lastly, note the level
ordering of the exotic isoscalar $u\overline{u}g$ and
$s\overline{s}g$ spectra are the same, $0^{--}$, $1^{-+}$,
$1^{-+}$, $1^{-+}$ and $3^{-+}$,  but slightly different than the
exotic $c\overline{c}g$ system, where the $0^{--}$ and lowest
$1^{-+}$ are degenerate. This is a consequence of the enhanced
charmonium self-energy from the hyperfine interaction.

Finally, we discussed both low and high energy scenarios for
observing hybrid mesons.  For low energy studies involving light
quark systems, dimensional counting rules predict specific
power-law behaviors for distinguishing between production of
conventional and hybrid mesons.  For high energy investigations of
heavy quark systems, the Frank-Condon principle provides a useful
constraint on the final state momentum distributions which should
assist experimentalist in identifying heavy hybrid systems.

In summary,  lattice, Flux Tube and our $H_{\rm eff}$ many-body
approach all predict similar hybrid spectra and that the lightest
$1^{-+}$ exotic hybrid meson mass is near 2 GeV.  This composite
model agreement indicates that the $\pi_1(1600)$ is not a hybrid
meson but has an alternative structure. If true and if the
$\pi_1(1600)$ exists, it is more likely a tetraquark system,
either a $(q \overline q )$ $(q \overline q )$ meson molecule or
an exotic $qq\bar{q} \bar{q}$ atom. Future work will apply our
model to light and heavy tetraquark systems including mixing with
hybrid and conventional meson states.  Three-body forces
\cite{Szczepaniak:2006nx} will also be examined.

\begin{acknowledgments}

Work supported in part by grants FPA 2004-02602, 2005-02327,
PR27/05-13955-BSCH (Spain) and U. S. DOE Grants DE-FG02-97ER41048
and DE-FG02-03ER41260.
\end{acknowledgments}



\begin{thebibliography}{99}

\bibitem{Alde}D. Alde \textit{et al.}, Phys. Lett. B \textbf{205}, 397 (1988).
\bibitem{E852-1}D. R. Thompson \textit{et al.} (E852 Collaboration), Phys. Rev. Lett. \textbf{79}, 1630 (1997).
\bibitem{E852-2}S. U. Chung \textit{et al.} (E852 Collaboration), Phys. Rev. D \textbf{60}, 092001 (1999).
\bibitem{E852-3}G. S. Adams \textit{et al.} (E852 Collaboration), Phys. Rev. Lett \textbf{81}, 5760 (1998).
\bibitem{E852-4}S. U. Chung \textit{et al.} (E852 Collaboration), Phys. Rev. D \textbf{65}, 072001 (2002).
\bibitem{ds}A. R. Dzierba {\it et al.}, Phys. Rev. D {\bf 73}, 072001 (2006).

\bibitem{Amelin}D. V. Amelin \textit{et al.} (VES Collaboration), Phys. Lett. B \textbf{356}, 595 (1995).
\bibitem{Zaitsev}A. Zaitsev, AIP Conf. Proc. \textbf{432}, 461 (1998).
\bibitem{Donnachie}A. Donnachie and Yu. S. Kalashnikova, Phys. Rev. D \textbf{60}, 114011 (1999).
\bibitem{Karch}K. Karch \textit{et al.} (Crystal Ball Collaboration), Z. Phys. C \textbf{54}, 33 (1992).
\bibitem{Adomeit}J. Adomeit \textit{et al.} (Crystal Barrel Collaboration), Z. Phys. C \textbf{71}, 227 (1996).
\bibitem{Barberis}D. Barberis \textit{et al.} (WA102 Collaboration), Phys. Lett. B \textbf{413}, 217 (1997).
\bibitem{Kalashnikova:2001ke}
  Yu.~S.~Kalashnikova,
  Nucl.\ Phys.  {\bf A689}, 49 (2001).
\bibitem{Buisseret:2006sz}
  F.~Buisseret and V.~Mathieu,
  arXiv:hep-ph/0607083.
\bibitem{Bernard1} C. Bernard \textit{et al.}, Phys. Rev. D \textbf{56}, 7039 (1997).
\bibitem{Bernard2} C. Bernard \textit{et al.}, Nucl. Phys.  (Proc. Suppl.) \textbf{B73}, 264 (1999).
\bibitem{Lacock} P. Lacock and K. Schilling, Nucl. Phys.  (Proc. Suppl.) \textbf{B73}, 261 (1999).
\bibitem{Hedditch}J. N. Hedditch \textit{et al.}, Phys. Rev. D \textbf{72}, 114507 (2005).
\bibitem{Luo} X. Q. Luo and Z. H. Mei, Nucl. Phys.  (Proc. Suppl.) \textbf{B119}, 263 (2003).
\bibitem{Barnes}T. Barnes, F. E. Close and E. S. Swanson, Phys. Rev. D \textbf{52}, 5242 (1995).
\bibitem{Close}F. E. Close and P. R. Page, Nucl. Phys.  \textbf{B443}, 233 (1995).
\bibitem{Katja}K. Waidelich, Diploma Thesis, North Carolina State University (2001).
\bibitem{Bag}T. Barnes, Ph.D. Thesis, Caltech (1977); Nucl. Phys.  \textbf{B158}, 171 (1979);
T. Barnes and F. Close, Phys. Lett. B \textbf{116}, 365 (1982); M.
Chanowitz and S. Sharpe, Nucl. Phys. \textbf{B222}, 211 (1983); T.
Barnes \textit{et al.}, Nucl. Phys. \textbf{B224}, 241 (1983); M.
Flensburg \textit{et al.}, Z. Phys. C \textbf{22}, 293 (1984); P.
Hasenfratz \textit{et al.}, Phys. Lett. B \textbf{95}, 299 (1980).
\bibitem{Iddir}F. Iddir and L. Semlala, arXiv:hep-ph/0511086.
\bibitem{Liu} Y. Liu and X. Q. Luo, Phys. Rev. D {\bf 73}, 054510 (2006).\bibitem{Griffiths}L. A. Griffiths, C. Michael and P. E. L. Rakow, Phys. Lett. B \textbf{129}, 351 (1983).
\bibitem{Perantonis}S. Perantonis and C. Michael, Nucl. Phys.  \textbf{B347}, 854 (1990).
\bibitem{LC2}
F. J. Llanes-Estrada and S. R. Cotanch, Nucl. Phys.
{\bf A697}, 303 (2002).
\bibitem{LCSS}
F. J. Llanes-Estrada, S. R. Cotanch, A. P. Szczepaniak and E. S. Swanson, Phys.
Rev. C  {\bf 70}, 035202 (2004).
\bibitem{LBC}
F. J. Llanes-Estrada, P. Bicudo and S. R. Cotanch, Phys.
Rev. Lett. {\bf 96}, 081601 (2006).
\bibitem{LC1}
F. J. Llanes-Estrada and S. R. Cotanch, Phys.
Rev. Lett. {\bf 84}, 1102 (2000).
\bibitem{LChybrid}
F. J. Llanes-Estrada and S. R. Cotanch, Phys. Lett. B
{\bf 504}, 15 (2001).
\bibitem{T-D-Lee}T. D. Lee, {\it Particle Physics and Introduction to Field Theory}  (Harwood Academic Publishers, New York, 1990).
\bibitem{SS}
A. P. Szczepaniak and E. S. Swanson, Phys.
Rev. D  {\bf 65}, 0252012 (2002).

\bibitem{Vegas}  G. P. Lepage, Journal of Comput. Phys. \textbf{27}, 192 (1978);
         Cornell University Report  CLNS 80-447, 1980 (unpublished).
\bibitem{PDG}S. Eidelman \textit{et al.}, Phys. Lett. B \textbf{592}, 1 (2004).
\bibitem{Brodsky}
S. J. Brodsky and G. R. Farrar, Phys. Rev. Lett. {\bf 31}, 1153 (1973); Phys. Rev. D {\bf 11}, 1309 (1975).
\bibitem{Gunionplb79}
J. F. Gunion, Phys. Lett. B {\bf 88}, 150 (1979).
\bibitem{Bodwinprl}
G. T. Bodwin, E. Braaten and J. Lee, Phys.Rev. D {\bf 72},  014004 (2005).
\bibitem{Szczepaniak:2006nx}
  A.~P.~Szczepaniak and P.~Krupinski,
  Phys.\ Rev.\ D {\bf 73}, 116002 (2006).

\end{thebibliography}
\end{document}